# End-to-End Mispronunciation Detection and Diagnosis From Raw Waveforms

Bi-Cheng Yan, Berlin Chen
*Dept. of Computer Science and Information Engineering*
*National Normal Taiwan University*
Taipei, Taiwan
{80847001s, Berlin}@ntnu.edu.tw

*Abstract*—Mispronunciation detection and diagnosis (MDD) is designed to identify pronunciation errors and provide instructive feedback to guide non-native language learners, which is a core component in computer-assisted pronunciation training (CAPT) systems. However, MDD often suffers from the data-sparsity problem due to that collecting non-native data and the associated annotations is time-consuming and labor-intensive. To address this issue, we explore a fully end-to-end (E2E) neural model for MDD, which processes learners' speech directly based on raw waveforms. Compared to conventional hand-crafted acoustic features, raw waveforms retain more acoustic phenomena and potentially can help neural networks discover better and more customized representations. To this end, our MDD model adopts a co-called SincNet module to take input a raw waveform and covert it to a suitable vector representation sequence. SincNet employs the cardinal sine (sinc) function to implement learnable bandpass filters, drawing inspiration from the convolutional neural network (CNN). By comparison to CNN, SincNet has fewer parameters and is more amenable to human interpretation. Extensive experiments are conducted on the L2-ARCTIC dataset, which is a publicly-available non-native English speech corpus compiled for research on CAPT. We find that the sinc filters of SincNet can be adapted quickly for non-native language learners of different nationalities. Furthermore, our model can achieve comparable mispronunciation detection performance in relation to state-of-the-art E2E MDD models that take input the standard hand-crafted acoustic features. Besides that, our model also provides considerable improvements on phone error rate (PER) and diagnosis accuracy.

## I. Introduction

With accelerating globalization, more and more people are willing or required to learn second languages (L2). Developments of computer-assisted pronunciation training (CAPT) systems open up new possibilities to enable L2 learners to practice L2 pronunciation skills in an effective and stress-free manner [1][2]. As an integral component of a CAPT system, mispronunciation detection and diagnosis (MDD) manages to provide different kinds of information, such as pronunciation scores [5] and diagnosis feedback [6], to guide non-native language learners to practice their pronunciations. Pronunciation scores reflect how similar an L2 learner's pronunciation is to that of native speakers. In practice, pronunciation scoring based methods detect errors using confidence measures that are derived from the automatic speech recognizer (ASR), e.g., phone durations, phone posterior probability scores and segment duration scores [7][8]. In order to obtain informative diagnosis feedback, the extended recognition network (ERN) method augments the decoding network of ASR with phonological rules. By comparison between an ASR output and the corresponding text prompt, ERN can readily offer appropriate diagnosis feedback on mispronunciations [9][10]. However, it is practically difficult to enumerate and include sufficient phonological rules into the decoding network for different L1-L2 language pairs. Furthermore, inclusion of too many phonological rules would incur ASR accuracy drop and in turn lead to poor MD performance. Apart from the above, a common thought is that we can evaluate learners' pronunciations based on free phone recognition. An MDD system is thus trained to directly recognize the possible sequence of phones pronounced by a non-native learner, which can be compared to canonical phone sequence of a given prompt [11][12][13]. Problems facing this approach include scarce annotated training data for deviated pronunciations and the variability in how different speakers articulate each phone. Alternatively, we can conduct MDD by recognizing articulatory features instead of phone symbols [14][15] or employing a multi-task learning strategy to leverage additionally speech data compiled from L2 speakers' native languages [16].

Recently, several efforts have been made to utilize deep neural networks to learn complex and abstract representations directly based on speech waveforms for ASR [17][18]. Compared with using traditional hand-crafted acoustic features, raw waveform modelling allows the incorporation of more information cues, such as the phase spectrum information that is typically ignored in Mel-scale frequency cepstral coefficients (MFCCs) derived from Fourier transform magnitude-based spectra. Furthermore, such human-engineered features, in fact, are originally designed from perceptual evidence and there is no guarantee that such representations are really suitable for all speech-related tasks. Directly processing the raw waveform allows neural network based methods to learn low-level acoustic representations that are possibly more tailored to a specific task. Palaz *et al.* [17] investigated the usefulness of raw waveform-based models on the TIMIT phone recognition task, which showed that CNNs could deliver superior performance over fully connected networks. Ravanelli *et al.* [18] introduced a parametrized sinc functions in replace of the convolution operators of CNNs and proposed the so-called SincNet architecture, which has shown promising results on various speaker identification, verification and phone recognition tasks. In the context of MDD, Yang *et al.* [15] proposed an unsupervised approach and argued that the incorporation of large-scale unlabeled native speech data in the training stage can overcome the data sparsity problem for MDD. The authors employed a contrastive predictive coding (CPC) model trained with language-adversarial training criteria to align the feature distributions between the L1 and L2 speech datasets, yielding accent-invariant speech representations for MDD.

In this paper, we develop a fully end-to-end (E2E) neural model architecture that streamlines the MDD process by directly taking input the waveform signals of L2 learners' speech. Specifically, an E2E free phone recognition system is adopted to generate the possible phone sequence uttered by an L2 learner, which is instantiated by a hybrid connectionist temporal classification/attention (CTC-ATT) model [19]. Further, we insert a SincNet (or CNN) module prior to the encoder of CTC-ATT model, which aims to covert a raw waveform signal to a suitable vector representation sequence. The overall model architecture is depicted in Fig. 1. By comparison to CNN, SincNet has fewer parameters and is more amenable to human interpretation. Empirical evaluations conducted on the L2-ARCTIC dataset demonstrate that our proposed SincNet-based E2E MDD model significantly reduces the phone error rate (PER). Furthermore, our best model can achieve comparable mispronunciation detection performance in relation to these state-of-the-art models.

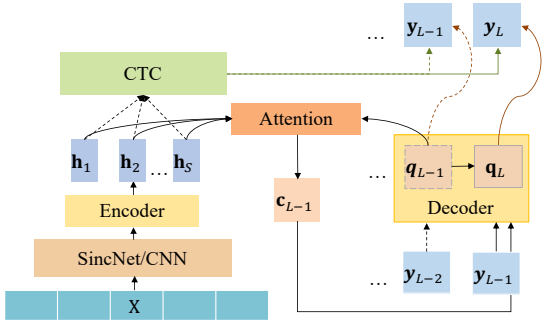

Fig. 1. Mispronunciation detection and diagnostic with a hybrid CTC-Attention ASR architecture from raw waveforms.

## II. SINCNET-BASED E2E MDD MODEL

### A. Hybrid CTC-ATT ASR Model

The basic hybrid CTC-ATT MDD model typically consists of four modules, as depicted in Fig. 1: 1) an encoder module that is shared across CTC and the attention-based model. The encoder module extracts $S$-length high-level encoder state sequence $H = (\mathbf{h}_1, ..., \mathbf{h}_S)$ from a $T$-length acoustic features $X = (\mathbf{x}_1, ..., \mathbf{x}_T)$ through a stack of convolutional and/or recurrent networks, where $S \leq T$ is due to downsampling; 2) an attention module that calculates a fixed-length context vector $\mathbf{c}_l$ by summarizing the output of the encoder module at each output step for $l \in [1, ..., L]$, finding out relevant parts of the encoder state sequence to be attended for predicting an output phone symbol $y_l$, where the output symbol sequence $\mathbf{y} = (y_1, ..., y_L)$ belongs to a canonical phone set $\mathcal{U}$; 3) Given the context vector $\mathbf{c}_l$ and the history of partial diagnostic results $y_{1:l-1}$, a decoder module updates its hidden state $\mathbf{q}_l$ autoregressively and estimates the next phone symbol $y_l$; 4) The CTC module offers another diagnostic results based on the frame-level alignment between the input sequence X and the canonical phone symbol sequences $\mathbf{y}$ by introducing a special <blank> token. It can substantially reduce irregular alignments during the training and test phases.

The training objective function of the hybrid CTC-ATT model is to maximize a logarithmic linear combination of the posterior probabilities predicted by CTC and the attention-based model, i.e., $p_{CTC}(\mathbf{y}|X)$ and $p_{att}(\mathbf{y}|X)$:

$$\mathcal{L} = \alpha \log p_{CTC}(\mathbf{y}|X) + (1-\alpha)\log p_{att}(\mathbf{y}|X), \quad (1)$$

$$p_{CTC}(\mathbf{y}|X) = \sum_{z} p(\mathbf{y}|\mathbf{z}, X)p(\mathbf{z}|X), \quad (2)$$

$$\approx \sum_{z} p(\mathbf{y}|\mathbf{z})p(\mathbf{z}|X), \quad (3)$$

$$p_{att}(\mathbf{y}|X) = \prod_{l=1}^{L} p(y_l|y_{1:l-1}, X). \quad (4)$$

where the frame-wise latent variable sequences $\mathbf{z}$ belongs to a canonical phone set $\mathcal{U}$ augmented with and the additional <blank> label, which facilitates CTC to enforce a monotonic behavior of phone-level alignments. Eq. (3) is a neat formulation of the CTC model, which is derived from the assumption that the symbol translation model $p(\mathbf{y}|\mathbf{z})$ is conditionally independent of the input sequence X [20]. The linear combination weight $\alpha$ in Eq. (1) is a hyperparameter used to linearly interpolate the two posterior probabilities. In our experiments, the $\alpha$ is set equal to 0.5.

The attention-based model can be expressed by

$$H = Encoder(X), \quad (5)$$

$$\boldsymbol{\alpha}_{l,s} = Attention(\mathbf{q}_l, \mathbf{h}_s), \quad (6)$$

$$\mathbf{c}_l = \sum_{s} \boldsymbol{\alpha}_{l,s} \mathbf{h}_s, \quad (7)$$

$$p(y_l|y_{1:l-1}, H) = Decoder(\mathbf{q}_{l-1}, [y_{l-1}; \mathbf{c}_{l-1}]), \quad (8)$$

In order to constrain the range of the attention operations, we choose the location-aware attention [19] as the attention mechanism in our experiments, as indicated in Eq. (7).

### B. The SincNet Architecture

SincNet [18] is a parametric counterpart used to replace the convolution operations of CNN. Each impulse response of SincNet's filters is a subtraction of two cardinal sine (sinc) functions, resulting in an ideal bandpass filter. The standard CNN perform a set of time-domain convolution operations on the input with some finite impulse response filters defined as:

$$s[t] = x[t] * h[t] = \sum_{l=0}^{L-1} x[l] \cdot h[t-l]. \quad (9)$$

where $x[t]$ is a chunk of the speech signal, $h[t]$ is a filter of $L$-length, and $s[t]$ is the filtered output. In this case, all the elements of $h[\cdot]$ are learnable parameters (i.e., all the $L$ elements of each filter are learned from data). SincNet proposes to replace filters $h$ with a sinc function $g$ that only depends on two variables: low and high cut-off frequencies. A filter in SincNet with impulse response $g[t, f_1, f_2]$ and frequency response $G[f, f_1, f_2]$ can be respectively express by:

$$g[t, f_1, f_2] = 2f_2 sinc(2\pi f_2 t) - 2f_1 sinc(2\pi f_1 t), \quad (10)$$

$$G[f, f_1, f_2] = \Pi\left(\frac{f}{2f_2}\right) - \Pi\left(\frac{f}{2f_1}\right). \quad (11)$$

where $sinc(x) = \frac{\sin(x)}{x}$, $f_1$ and $f_2$ are two learnable parameters that describe low and high cutoff frequencies, $f$ is a frequency index, and $\Pi(\cdot)$ is a rectangular bandpass filter function in the magnitude frequency domain.

TABLE I. STATISTICS OF THE EXPERIMENTAL SPEECH CORPORA.

| Subsets | #Spks | #Utters | #Hours | #Phones | #Marked errors |
|---|---|---|---|---|---|
| Train | 477 | 6237 | 8.81 | 234K | 13K (0.56%) |
| Test | 6 | 900 | 0.87 | 15K | 4K (13.65%) |

TABLE II. CONFUSION MATRIX OF MISPRONUNCIATION DETECTION AND DIAGNOSIS.

| Total condition | | Ground truth | |
|---|---|---|---|
| | | CP | MP |
| Model prediction | CP | True positive (TP) | False positive (FP) |
| | MP | False negative (FN) | True negative (TN=CD+DE) |

## III. EXPERIMENTAL SETTINGS AND PERFORMANCE EVALUATION METRICS

### A. Speech Corpora and Model Configuration

We carried out experiments using the L2-ARCTIC [21] and TIMIT corpus [22] for MDD tasks. The L2-ARCTIC dataset is a publicly-available non-native English speech corpus compiled for research on CAPT, accent conversion, and others. It contains correctly pronounced utterances and mispronounced utterances of 24 non-native speakers (12 males and 12 females), whose L1 languages include Hindi, Korean, Mandarin, Spanish, Arabic and Vietnamese. Apart from that, a suitable quantities of native (L1) English speech datasets compiled from the TIMIT corpus (composed of 630 speakers) was used to bootstrap the training of the various E2E MDD models. To unify the phone sets of these two corpora, we followed the definition of the CMU pronunciation dictionary to obtain an inventory of 39 canonical phones. Next, we divided these two corpora into training, development and test sets, respectively; in particular, the setting of the mispronunciation detection experiments on L2-ARCTIC followed the recipe provided by [13]. Table I summarizes detail statistics of these speech datasets.

Our baseline E2E MDD models built on the hybrid CTC-ATT model. The encoder network is composed of the VGG-based deep CNN component plus a bidirectional LSTM component with 1024 hidden units [19], which takes input the hand-crafted acoustic features, such as Mel-filterbank outputs (FBANK) or MFCCs. The decoder network consists of two-layer unidirectional-LSTM with 1024 cells. As to the hand-crafted acoustic features, FBANK is 80-dimensional while MFCCs 40-dimensional. Both of them were extracted from waveform signals with a hop size of 10 ms and a window size of 25 ms, and further normalized with the global mean and variance. When taking input raw waveform signals alternatively, the SincNet module is tacked in front of the encoder network. SincNet module was based on the configuration suggested in [18], which is made of an array of parametrized sinc-functions in the first layer, followed by two one-dimensional convolutional layers. Further, each layer of SincNet has 80, 128 and 128 filters and kernel size are set to be 251, 3 and 3, respectively.

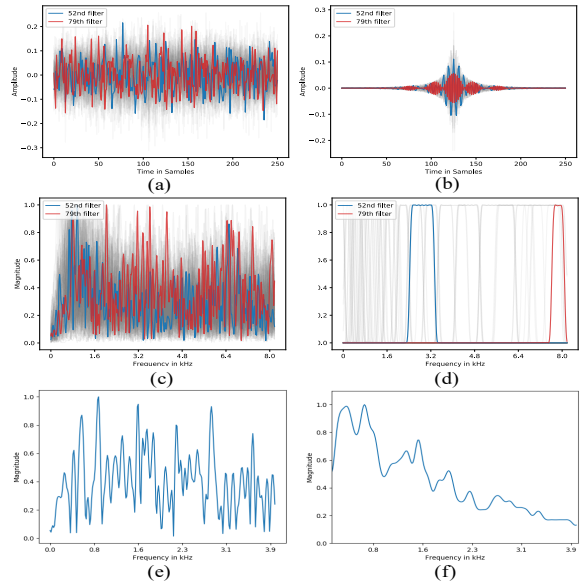

Fig. 2. Visualization of learned filters in the conventional CNN and SincNet. Two filters were randomly chosen and highlighted with different colors; the remaining filters were plotted with shade grey. (a) and (b) are impulse responses for CNN and SincNet; (c) and (d) are frequency responses for CNN and SincNet; and (e) and (f) show the normalized average frequency response for CNN and SincNet.

### B. Performance Evaluation Metrics

For the mispronunciation detection task, we follow the hierarchical evaluation structure adopted in [5], while the corresponding confusion matrix for four test conditions is illustrated in Table II. Based on the statistics accumulated from the four test conditions, we calculate the values of different metrics like recall, precision and the F-1 measure (the harmonic mean of the precision and recall), so as to evaluate the performance of mispronunciation detection. Those metrics are defined as follows:

$$\text{Precision} = \frac{\text{TN}}{\text{TN} + \text{FN}}, \quad (12)$$

$$\text{Recall} = \frac{\text{TN}}{\text{TN} + \text{FP}}, \quad (13)$$

$$\text{F-1} = 2\frac{\text{Precision} * \text{Recall}}{\text{Precision} + \text{Recall}}. \quad (14)$$

Furthermore, to calculate the diagnostic accuracy rate (DAR), we focus on the cases of TN and consider it as combination of diagnostic errors (DE) and correct diagnosis (CD). The accuracies of mispronunciation diagnosis rate (DAR) are calculated by:

$$\text{DAR} = \frac{\text{CD}}{\text{CD} + \text{DE}}. \quad (15)$$

## IV. EXPERIMENTAL RESULTS

### A. Interpretation of the learned filters

At the outset, we analyze the dynamic properties of the learned filters in the first layer of the CNN (or SincNet) module, respectively. Fig. 2 illustrates the impulse and frequency responses of the learned filters, which were trained on the correctly-pronounced training utterances of L2 learners. In the

TABLE III. %PER FOR CORRECT PORNUNCIATION UTTERENCES WITH DIFFERENT FRONTEDN PROCESSING SCHEMES.

|  | MFCC | FBANK | CNN | SincNet |
|---|---|---|---|---|
| %PER | 9.25 | 8.45 | 6.44 | 5.50 |

TABLE IV. MISPRONUNCIATION DETECTION RESULTS FOR DIFFERENT MODELS.

| Model | Input Feature | %Recall | %Precision | %F1 |
|---|---|---|---|---|
| GOP | FBANK | 52.88 | 35.42 | 42.42 |
| CTC-ATT | MFCC | 53.54 | 53.64 | 53.59 |
| CTC-ATT | FBANK | 52.43 | 55.31 | 53.83 |
| CTC-ATT +CNN | RAW | 47.60 | 55.15 | 51.10 |
| CTC-ATT +SincNet | RAW | 50.09 | 55.31 | 52.57 |

TABLE V. DETAILS OF MISPRONUNCIATION DETECTION AND DIAGNOSIS RESULTS FOR DIFFERENT MODELS.

| Models | Correct pronunciations | | Mispronunciations | | %DAR |
|---|---|---|---|---|---|
|  | %TP | %FN | %TN | %FP |  |
| Leung et al.* [11] | 67.81 | 32.19 | 65.04 | 32.96 | 32.10 |
| CTC-ATT† | 89.79 | 10.21 | 52.43 | 47.57 | 59.84 |
| CTC-ATT +CNN | 90.67 | 9.33 | 47.60 | 52.40 | 62.08 |
| CTC-ATT +SincNet | 90.25 | 9.75 | 50.09 | 49.91 | 60.96 |

Note: *We reproduced the model architecture in the framework of *Leung et al* by adopting FBANK features for the CNN-RNN-CTC model, so there may exists some slight differences with [11]. †We report on the CTC-ATT model that takes input the FBANK features.

first raw, we depict the learned filters in the time domain. The learned filters of CNN are a set of filters with $L$ parameters (i.e., $L = 251$). Thus, the impulse response for CNN seemingly learn to represent temporal property of a waveform and its corresponding frequency response looks without particular patterns, which is uniformly distributed in low and high frequency bins (*cf*. Fig. 2 (a) and (c)). SincNet, instead making use of a set of sinc functions which are designed to implement rectangular bandpass filters, demonstrate regularity in the frequency domain. Looking at Fig. 2 (d), we can find that most the filters of SincNet lie on frequencies below 2,000 Hz. This implies that SincNet learns to reflect the properties of human's auditory system. We then turn to examine the normalized average frequency responses of CNN and SincNet. Fig. 2 (e) and (f) depict the normalized average frequency responses of the filters learned by CNN and SincNet, respectively. There are several obvious peaks standing out in the plot of SincNet. The observation to some extent is consistent with [18], which mentioned that the filters of SincNet capture mainly the property of pitch regions (the average pitch is 133 Hz for a male and 234 for a female) and cover various English vowels regions (the first and second formant of English vowels approximately located at 500 Hz and 1,300 Hz, respectively).

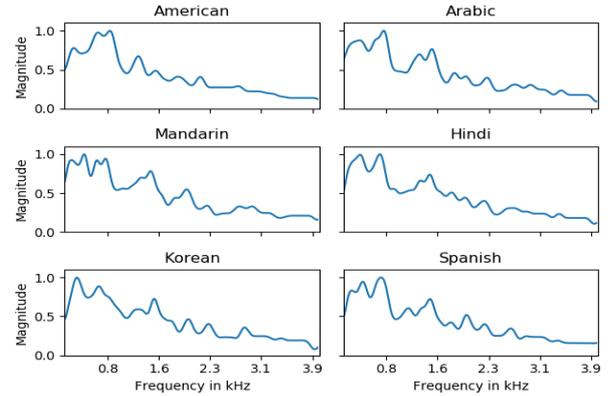

Fig. 3. Normalized average frequency response for different L1-dependent pronunciation distributions extracted by SincNet.

Next, in order to examine the properties of SincNet filters learned for the L1-dependent pronunciation distributions of L2 learners with different nationalities, we use all of training data to build a multilingual SincNet module and then transfer it with the L1 speech data of L2 learners to obtain their respective L1-dependent SincNet module. The corresponding normalized average frequency response be demonstrated in Fig. 3. As can be observed, all of the L1-dependent filters operate in frequency bins below 2,000 Hz, which is consistent with the perceptual scales that take inspiration from the human auditory system. This implies that the learned filters essentially were L1-dependent and selective in processing those spectral components. Further, each filter seemingly learns to capture different English vowels, in terms of their first or second formants. For example, the filters of American, Arabic, Hindi and Spanish clearly manifest the vowel /a/ according to its second formant (the second formant of vowel /a/ centers around 1,100Hz). Furthermore, the filters of Mandarin and Korean tend to highlight the frequencies below 800 Hz. We argue that these filters learn to represent the vowel /u/ or /ɔ/ (the second formant of vowel /u/ and /ɔ/ are below 800 Hz). It is worth mention that the L1 speech dataset for each nationality of the L2 learners are approximate 4 hours. This indicates that the filters involved in SincNet can be adapted quickly for different non-native language learners.

### B. Comparison of Phone Recognition Performation

We report the phone error rate (PER) for CTC-ATT model trained with raw waveform modeling (CNN, SincNet) or traditional hand-crafted acoustic features (MFCC, FBANK). In this experiment, the testing utterances were subset of testing data set, which adopt correct part of pronounced utterances only. The results are summarized in Table III. We can observe that both of raw waveform models significantly outperform the traditional features models. SincNet can lead to lowest PER, at least have 3% absolute PER reduction while compared to the traditional hand-crafted acoustic features. Next. The performance degraded when SincNet was swapped with a standard CNN.

### C. Performance of Mispronunciation Detection

We assess the performance levels for various different input formats for E2E MDD, i.e., FBANK features, MFCCs, and raw waveform signals, with respect to mispronunciation detection. We also report the pronunciation scoring based MDD method, namely the GOP method building on the hybrid DNN-HMM ASR model. Specifically, the DNN component of GOP is a 5-layer time-delay neural network (TDNN) with 1,280 neurons in each layer. The corresponding results are shown in Table IV. As can be seen, the E2E ASR based methods clearly surpass the GOP-based method, with at least 10% absolute improvements in terms of the F-1 measure. This implies that the free phone recognition method can boost the performance on mispronunciation detection. Next, as traditional hand-crafted acoustic features are taken as the input to the encoder network of the E2E MDD model, we can find that the performance of FBANK is on par with MFCCs. FBANK model stands out in performance when precision is used as the evaluation metric, whereas the situation is reversed when recall is used as the metric. Finally, we compare the performance of CNN and SincNet, both of which consume raw waveform signals to the encoder network. As can be seen from Table IV, SincNet is slightly better than CNN by 1.47%, in terms of the F-1 measure. Interestingly, SincNet achieves comparable MD results with the E2E MDD model that takes input the FBANK features.

*D. Performance of Mispronunciation Detection Diagnosis*

In the third of set of experiments, we turn to evaluating the mispronunciation diagnosis performance of different E2E MDD models. The corresponding results are shown in Table V, where the true positive rate (TP) and true negative rate (TN) are two important criteria for evaluating the performance of CAPT systems. From this Table V, we can observe that our CTC-ATT model can boots CNN-RNN-CTC model proposed from Leung *et al.*, by consulting the diagnosis results produced by the output of the attention-based decoder. Second, when the CTC-ATT model takes input raw waveform signals with either CNN or SincNet (i.e., the streamlined E2E model), it can slightly perform better than the CTC-ATT model takes input FBANK features (i.e., the cascaded E2E model) in terms of true positive rate, whereas the situation is reversed when consider the true negative rate. It implies that the streamlined E2E model can learns to be more discriminative on correct pronunciation detection. Finally, we turn to investigating the mispronunciation diagnosis rate (DAR). As can be seen, using CNN to extract acoustic features from raw waveform signals can achieve the best performance. Besides that, the DAR of using SincNet to extract acoustic features from raw waveform signals also better than the baseline E2E model that takes input FBANK features.

## V. Conclusion

In this paper, we have designed and developed a fully end-to-end (E2E) neural model architecture that streamlines the MDD process by taking input waveform signals uttered by L2 learners directly. Promising results have been obtained through a series of empirical experiments conducted on the L2-ARCTIC benchmark dataset. As to future work, we will try to investigate contrastive predictive coding (CPC) model and expect to solve the data sparsity problem through leveraging datasets of native speakers. Furthermore, we also intended to study the suprasegmental-level phenomena for CAPT, such as intonation, accent pitch and rhythm.